# AN EVIDENCE OF LINK BETWEEN DEFAULT AND LOSSES OF BANK LOANS FROM THE MODELING OF COMPETING RISKS

**Mauro R. Oliveira[1], Francisco Louzada[2]**
[1]*Department of Statistics, Federal University of Sao Carlos, Brazil, (UFSCar)*
[2]*Department of Applied Mathematics & Statistics, University of Sao Paulo, Brazil (USP)*

**Abstract**
In this paper, we propose a method that provides a useful technique to compare relationship between risks involved that takes customer become defaulter and debt collection process that might make this defaulter recovered. Through estimation of competitive risks that lead to realization of the event of interest, we showed that there is a significant relation between the intensity of default and losses from defaulted loans in collection processes. To reach this goal, we investigate a competing risks model applied to whole credit risk cycle into a bank loans portfolio. We estimated competing causes related to occurrence of default, thereafter, comparing it with estimated competing causes that lead loans to write-off condition. In context of modeling competing risks, we used a specification of Poisson distribution for numbers from competing causes and Weibull distribution for failures times. The likelihood maximum estimation is used to parameters estimation and the model is applied to a real data of personal loans.

## 1 INTRODUCTION

Financial lending market in emerging countries has become more competitive due, among others reasons, increasingly presence of international banks operating abroad, seeking to increase their customer bases in relentless pursuit of growing sales and revenues. Particularly in Brazil, it has become challenger in recent years due initiative of public banks, which since end of 2011, have been promoting a drastic reduction in interest rates charged, and consequently made also all other players reduce their interest rates.
In this scenario, to achieve an optimal relationship between risk and return is necessary keep on changing efficient policies of credit risk management. Thus, to improve their ability to grant and generate profit, it is needed that banks have capability to forecast risks of their activities, for instance, default rates and losses arising from defaulted loans. The major intention is provide banks to make right decision about if, when, and how much they can reduce interest rates charged, in order to continue operating with competitiveness, profitability and soundness.
One risk analyst role is identify risk patterns in past behavior, and thereafter, he has chance to anticipate trends in levels of risk and compare them to levels that bankers are comfortable to take. We seek to answer in this article if there is a relationship between an increased risk of the portfolio, i.e., in relation to occurrence of default, with an increased in losses in defaulters loans. This positive relationship implies that in times of economic recession, where default rates tend to be higher, we would have less effectiveness also in the process of recovery on defaulted loans.

In a bank loans portfolio, we are faced with a large amount of data from loans that are still in progress and contracts that may have been finalized. Particularly for this work, we have available a data set of defaulted contracts awarded between 2007 and 2011. They are divided into 5 seasons according to grant year, and followed up until time when customer has become defaulter, i.e., within more than 90 days without paying back obligations. Our database has also information about the collection process, i.e., 24 months monitoring all defaulted loans when it goes into recovery and collection process. If not successful in the collection process, contract is considered write-off. The recovery process in question considers a loan recovered if 100 % of the contract amount in arrears was recovered. Contracts with lower recovery rates were not considered in this study.

Therefore, the application data set has information of time until occurrence of default, and the time until to recovery, or not recovery. Because characteristics of data available, it is natural to apply statistical methods developed from survival analysis. These techniques are commonly used when we have information about time until occurrence of a specific event.

The development and application of this theory to real data sets are discussed extensively in academic literature, particularly in medicine area, where, for instance, it is studied survival of patients undergoing different types of treatments and drugs. We suggest the reader to references in texts Maller and Zhou (1996) and Ibrahim et al. (2005). The modeling of competing risks is widespread in literature in articles such as Cooner et al. (2006), Cooner et al. (2007), Xu et al. (2011). Among a wide number of papers, this subject have been aware due important in articles as Chen et al. (1999), Tsodikov et al. (2003) and Tournoud and Ecochard (2007).

The competing risks modeling is part of survival analysis framework and, so far by the authors' knowledge, has not been applied to model risk behavior of loan portfolio during credit cycle. We analysed, over a long period, behavior of latent risk that leads to both events of our interest: default and recovery. It is our main goal, and the reason we are using competing risks model is that it is possible to give a practical interpretation of the parameters obtained.

Finally, in hold of outcomes obtained from application of modeling competing risks, we compared estimated latent risks in process that leads to default with estimates in collection process. Our main result: we have found that there is a positive relationship between the intensity of defaults and losses given defaults during recovery processes.

The forthcoming sections are organized as follows. The model formulation and inference methods occupy the Section 2. An application to a real data set is developed in Section 3. Finally, Section 4 concludes with some general remarks.

## 2 MODEL FORMULATION

In scenario of competing risk modeling, a failure or specific event of interest is considered as a driven result of causes occurring simultaneously over time. Therefore, to design this modeling is assigned two statistical distributions: one to random variable "time to event", as well as, one random variable representing amounts of events risks competing. For a further study of subject, we suggest, among others, reader to consult Crowder (2001) and Pintilie (2006).

There is a feature of observed data that we must take into account for model: presence of individuals in population that are not observed event of interest. This suggests we should consider that number of competitive risks could also be zero. For these individuals, the literature has called them for cure, or non-susceptible to event, due applications of theory in medical setting.

Our data set is a large sample of contracts that have defaulted, then all individuals reached event of default. On other hand, we have a second view on this basis, where all contracts were followed until a period of 24 months and may or may not be recovered at the end. Hence, this data set has contracts that achieved event of interest (default) and may or not achieve event of interest (recovery).

Therefore, this section is divided into two parts. In the first, we show model to fit time to default and competitive risks that lead to default. In the second subsection, we outlined time until recovery modeling, where presents model reformulation-suiting presence of non-recovery.

We use in this paper the probability distributions most commonly used in literature for survival analysis and competing risk modeling. Let us assume that time to event follows the Weibull distribution and amounts of risk events follow the Poisson distribution.

From above consideration, we present in next two subsections formulation of competing risk models considering where all population reached event of interest, as well as, there is presence of non-susceptible in population.

## 2.1 Zero-Truncated Poisson-Weibull distribution for Defaulted loans data

This model formulation assumes that all the elements of the database are susceptible to the event of interest, and then exist at least one cause of risk for all elements, thereafter, the amount of competing risks for the event is greater than zero.

The model is outlined as follows and, further details about this model can be find in Bereta (2011).

Let $M$ be a random variable, denoting the number of risks related to the occurrence of an event of default.

We assume that M has a zero truncated Poisson distribution with a probability mass function given by

$$P_{zt}(M = m) = \frac{\theta^m}{m!\,(\exp(\theta) - 1)}, \quad \text{where } m = 1,2,\ldots, \text{ and } \theta > 0. \quad (1)$$

Let $T_i$, for $i = 1,2,\cdots,$ denote the time-to-default due to the i-th factor of risks, which are assumed independent of M, and has the Weibull probability density function given by

$$f(t) = \gamma \beta^\gamma t^{\gamma-1} \exp(-(\beta t)^\gamma), \quad \text{where } t \geq 0, \quad \gamma > 0, \quad \beta > 0. \quad (2)$$

In the latent risks scenario, the number of causes M and the lifetime $T_1$ associated with a particular cause are not observable (latent variables), but only the minimum lifetime among all causes is usually observed. Therefore, the component lifetime is defined as $Y = \min(T_1, T_2, \cdots, T_M)$.

According to Bereta (2011), following shows random variable Y has a probability density function given by

$$f_Y(t) = \frac{\theta \exp[\theta \exp(-(\beta t)^\gamma) - (\beta t)^\gamma]\beta^\gamma t^{\gamma-1}\gamma}{\exp(\theta) - 1}, \quad \text{where } t \geq 0, \theta > 0, \beta > 0. \quad (3)$$

The corresponding survival function is given by

$$S_Y(t) = \frac{\exp[\theta \exp(-\beta t^\gamma)] - 1}{\exp(\theta) - 1},$$
$$\text{where } t \geq 0, \theta > 0, \beta > 0. \qquad (4)$$

Note $\lim_{t \to +\infty} S_Y(t) = 0$, thus agreeing with our assumption that all individuals will present the event of interest at some point in time.

To estimating the parameters of a model, we use maximum-likelihood estimation (MLE), based is a method of Newton-Raphson yet implemented in the software R.

The MLE for model 3 is estimated from $L(\Theta|t) = \prod_1^n f_Y(\Theta|t)$, where $\Theta = (\theta, \gamma, \beta)$ is the parameter vector.

## 2.2 Poisson-Weibull distribution for Recovery loans data

The next formulation assumes there are elements could are not susceptible to event of recovery, then the amount of competing risks for event may also be zero. It is generally known by promotion time model and appeared in the literature on earlier works in Chen et al. (1999) and Yakovlev and Tsodikov (1996).

Hence, now M has assumed here a Poisson distribution with support starting in 0, with a probability mass function given by

$$P(M = m) = \frac{\theta^m \exp(-\theta)}{m!}, \qquad \text{where } m = 1,2,\ldots, \qquad \text{and } \theta > 0. \qquad (5)$$

In same way, let $T_i$, for $i = 1,2,\cdots$, denote the time-to-recovery due to the $i$-th factor of risks, which are assumed independent of M, and within all Weibull probability density function.

Again, the time of occurrence of interest event is defined as $Y = \min(T_1, T_2, \cdots, T_M)$, where $P(T_0 = \infty) = 1$ and $T_i$ follows Weibull distribution like in (2). Let $F$ denote Weibull cumulative distribution function (cdf).

The following, see Chen et al. (1999) and Yakovlev and Tsodikov (1996), shows that random variable Y has a probability density function given by

$$f_Y(t) = \theta f(t) \exp[-\theta(F(t))]. \qquad (6)$$

The corresponding survival function is given by

$$S_Y(t) = \exp[-\theta(F(t))]. \qquad (7)$$

According to our intention, this assumption admits possibility of a proportion of population not shows occurrence of event of interest. This fraction of non-susceptible is equals $\exp(-\theta)$, which is greater than zero. Not reach the event means that the contract was not recovered after the period of 24 months in process of collection. So, here we define the expected loss given default ($ELGD$), given by

$$ELGD = \exp(-\theta). \qquad (8)$$

To estimating parameters of a model, we use maximum-likelihood estimation (MLE), however, taken consideration that there are individuals do not observed recovered time. Hence, the MLE for model 6 is estimated from $L(\Theta|t) = \prod_1^n f_Y(\Theta|t)^{\delta_i} S_Y(\Theta|t)^{1-\delta_i}$, where $\Theta = (\theta, \gamma, \beta)$ is the parameter vector, and $\delta_i = 1$ if the loan is recovered and $\delta_i = 0$ if not.

## 3 APLICATION DATA

In model of competing risks is possible gives a practical view of the parameters obtained. The Poisson distribution, which adjusts the amount of risks that lead to the event of interest, has easy interpretation for comparison. The parameter $\theta$ represents the expected amount of the random variable. In Table 1 has shown a comparing of Poisson parameter in both cases: which leads to the default ($\theta$-default) and that leads to recovery ($\theta$-recovery).

The results were applications in contracts samples grouped by year of grant, comprising a base of around 20,000 contracts, with contracts awarded between 2006 and 2010, where the latest samples accompanied until 2013 for recovery purposes. For example, the parameter $\theta$-default of sample 2008 is related to contracts that were awarded in 2008 and became defaulter posteriorly.

Regarding the parameter $\theta$−recovery, were considered 5 samples of contracts that began in recovery process between 2007 and 2011. This explains why there is no parameters for default in 2011, as there is no recovery parameter for the year 2006.

**Table 1: Poisson-Weibull parameters**

| Year | $\theta$-default | $\theta$-recovery | Observed LGD | ELGD% |
|---|---|---|---|---|
| 2006 | 3.0820 | ... | ... | ... |
| 2007 | 1.9048 | 0.2861 | 78.107 | 78.057 |
| 2008 | 1.6734 | 0.3418 | 75.239 | 75.200 |
| 2009 | 1.1951 | 1.4607 | 54.505 | 54.514 |
| 2010 | 1.5646 | 3.0614 | 48.077 | 48.164 |
| 2011 | ... | 0.8044 | 60.385 | 60.386 |

We can clearly see that estimated in the process that leads to default risk was falling between 2006 and 2009, which increase again up to 2010. Conversely, the risks that lead to recovery has increased on period until 2010 and reflected a period where the bank loses less money. This trend of recovery turns down after the year 2010, exactly where the risk of default goes back up. These outcomes are ratified with estimated values of loss given default displayed in the last column of Table 1, which compared to the losses observed, we see that this model is quite accurate.

## 4 CONCLUSIONS

This paper presented an application of statistical methodology of competing risks that compared over time the latent number of factors that lead to the occurrence of events related to the cycle of credit from a bank loans.

The goal was achieved because we found a link where decreasing losses and its reverse sequence of fall has direct relation to the risk factors that lead to default. This implies that, in times of increased risk of default, the bank should provide that there would be a decline in the expectation of receiving loans in process of collection.

Finally, we call attention to this relationship that implies in times of economic recession, where default rates tend to be higher, we would have less efficacy with recovery of defaulted loans process, and therefore the bank should seek to improve their recovery and collection policies.
Acknowledgments: CNPq and FAPESP, Brazil sponsored the presented research.

## Appendix A

This appendix presents the tables of the MLE, standard error (SE), lower and upper limit of the confidence interval (LI and UI) and the p-value of the parameters of Zero-Truncated Poisson-Weibull (MLES) for Defaulted loans data, as well as, of the Poisson-Weibull for recovered loans data.

**Table 2: Summaries of Zero-Truncated Poisson-Weibull (MLES) for defaulted loans data**

| 2006 | | | | |
|---|---|---|---|---|
| Parameter | SE | LI | UI | p-value |
| $\gamma = 2.7082$ | 0.0119 | 2.6848 | 2.7315 | < 0.0001 |
| $\beta = 0.2223$ | 0.0239 | 0.1754 | 0.2691 | < 0.0001 |
| $\theta = 2.9149$ | 0.0667 | 2.7841 | 3.0457 | < 0.0001 |
| **2007** | | | | |
| Parameter | SE | LI | UI | p-value |
| $\gamma = 2.7189$ | 0.0144 | 2.6904 | 2.7473 | < 0.0001 |
| $\beta = 0.2269$ | 0.0221 | 0.1835 | 0.2704 | < 0.0001 |
| $\theta = 1.4644$ | 0.1364 | 1.1970 | 1.7318 | < 0.0001 |
| **2008** | | | | |
| Parameter | SE | LI | UI | p-value |
| $\gamma = 2.7973$ | 0.0170 | 2.7638 | 2.8307 | < 0.0001 |
| $\beta = 0.3315$ | 0.0203 | 0.2916 | 0.3713 | < 0.0001 |
| $\theta = 1.1361$ | 0.1645 | 0.8136 | 1.4586 | < 0.0001 |
| **2009** | | | | |
| Parameter | SE | LI | UI | p-value |
| $\gamma = 2.9223$ | 0.0228 | 2.8775 | 2.9672 | < 0.0001 |
| $\beta = 0.4400$ | 0.0207 | 0.3993 | 0.4807 | < 0.0001 |
| $\theta = 0.3677$ | 0.1552 | 0.0634 | 0.6721 | < 0.0001 |
| **2010** | | | | |
| Parameter | SE | LI | UI | p-value |
| $\gamma = 3.4099$ | 0.0182 | 3.3742 | 3.4456 | < 0.0001 |
| $\beta = 0.4495$ | 0.0176 | 0.4150 | 0.4841 | < 0.0001 |
| $\theta = 0.9736$ | 0.2032 | 0.5753 | 1.3719 | < 0.0001 |

**Table 3: Summaries of Zero-Truncated Poisson-Weibull (MLES) for recovered loans data**

| 2007 | | | | |
|---|---|---|---|---|
| Parameter | SE | LI | UI | p-value |
| $\gamma = 1.1687$ | 0.0521 | 1.0666 | 1.2708 | $< 0.0001$ |
| $\beta = 13.2155$ | 0.1029 | 13.0136 | 13.4173 | $< 0.0001$ |
| $\theta = 0.2861$ | 0.0664 | 0.1557 | 0.4164 | $< 0.0001$ |
| 2008 | | | | |
| Parameter | SE | LI | UI | p-value |
| $\gamma = 1.1430$ | 0.0299 | 1.0843 | 1.2016 | $< 0.0001$ |
| $\beta = 14.3917$ | 0.0730 | 14.2486 | 14.5348 | $< 0.0001$ |
| $\theta = 0.3418$ | 0.0438 | 0.2558 | 0.4278 | $< 0.0001$ |
| 2009 | | | | |
| Parameter | SE | LI | UI | p-value |
| $\gamma = 1.0082$ | 0.0207 | 0.9675 | 1.0489 | $< 0.0001$ |
| $\beta = 44.4901$ | 0.2382 | 44.0231 | 44.9571 | $< 0.0001$ |
| $\theta = 1.4607$ | 0.1707 | 1.1260 | 1.7954 | $< 0.0001$ |
| 2010 | | | | |
| Parameter | SE | LI | UI | p-value |
| $\gamma = 1.0647$ | 0.0203 | 1.0249 | 1.1045 | $< 0.0001$ |
| $\beta = 81.3458$ | 0.4332 | 80.4967 | 82.1950 | $< 0.0001$ |
| $\theta = 3.0614$ | 0.3752 | 2.3259 | 3.7970 | $< 0.0001$ |
| 2011 | | | | |
| Parameter | SE | LI | UI | p-value |
| $\gamma = 1.2417$ | 0.0196 | 1.2032 | 1.2802 | $< 0.0001$ |
| $\beta = 24.2691$ | 0.0964 | 24.0801 | 24.4581 | $< 0.0001$ |
| $\theta = 0.8044$ | 0.0719 | 0.6633 | 0.9455 | $< 0.0001$ |

## Appendix B

It is presented in the figures 1 and 2 the graphs of the functions of survivals estimated superimposed on the Kaplan-Meier estimates for the models presented in the section 2.1 and 2.2, respectively.

**Figure 1: Survival function estimated**

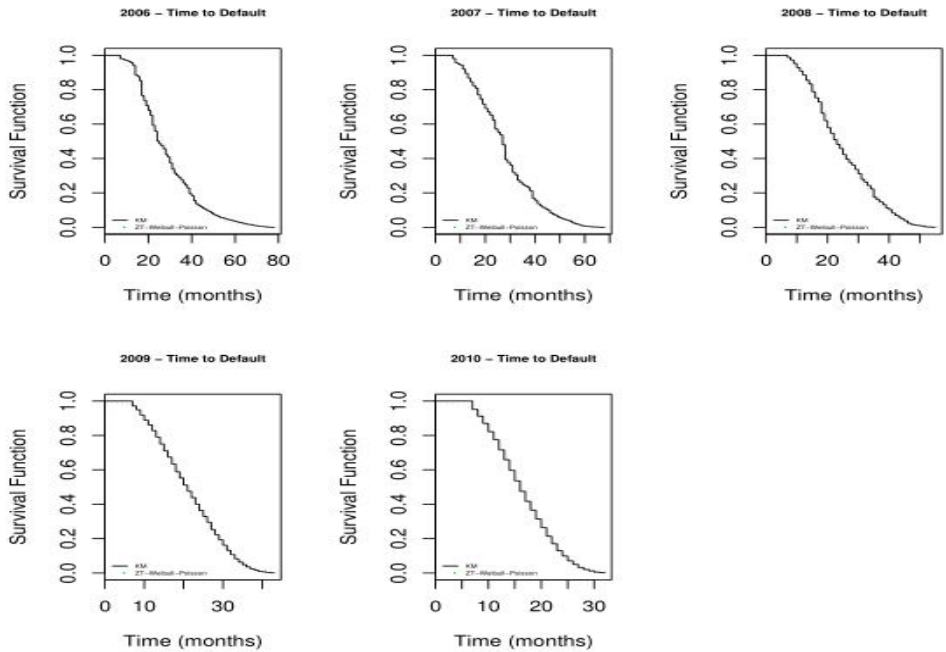

**Figure 3: Survival function estimated**

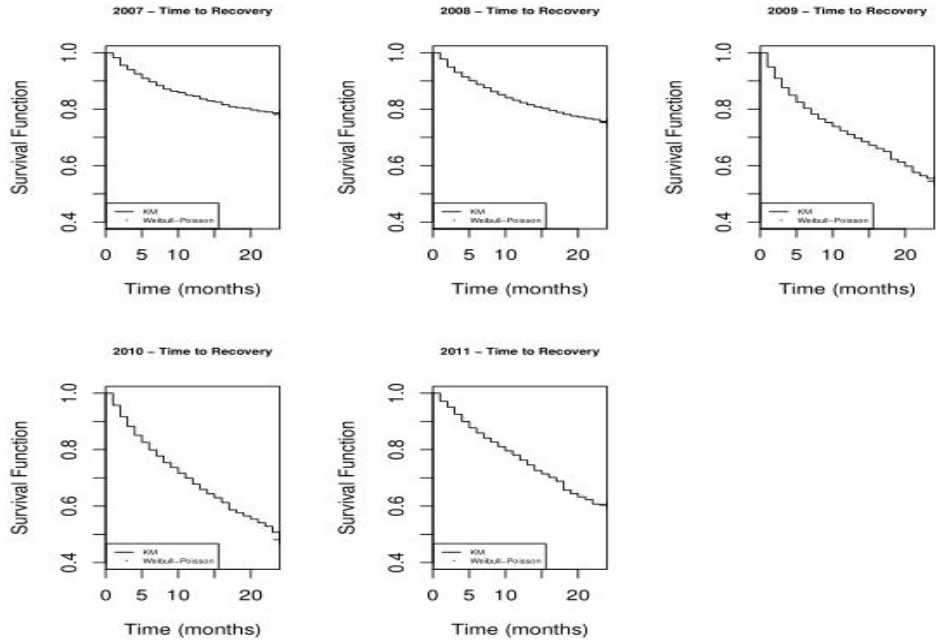